\documentclass[]{aa}  
\usepackage{color}
\usepackage{amssymb}

\usepackage{graphicx}
\usepackage{txfonts}
\usepackage[colorlinks, citecolor=blue]{hyperref}
\usepackage{mathtools}
\usepackage{units}
\begin{document} 

\title{Ongoing astrometric microlensing events of two nearby stars}
   
\author{J. Kl\"{u}ter\inst{1},
          U. Bastian\inst{1},
         M. Demleitner\inst{1},
        J. Wambsganss\inst{1,}\inst{2}}

   \institute{ Zentrum f\"{u}r Astronomie der Univ. Heidelberg, Astronomisches Rechen-Institut, M\"{o}nchhofstr. 12, 69120 Heidelberg, Germany\\
   \email{klueter@ari.uni-heidelberg.de}
        \and
   International Space Science Institute, Hallerstr. 6, 3012 Bern, Switzerland}

\authorrunning{J. Kl\"{u}ter et al.}
\titlerunning{Ongoing Astrometric Microlensing Events of two nearby Stars}

   \date{Received  May 21 , 2018; accepted June 14, 2018}
 
  \abstract
   {Astrometric microlensing is an excellent tool to determine the mass of stellar objects. 
By measuring the astrometric shift of a background source star in combination with precise predictions of its unlensed position and of the lens position, gravitational lensing allows to one determine the mass of the  lensing star with a precision of 1 percent, independently of any prior knowledge.
}
   {Making use of the recently published  Gaia Data Release 2 ({\it Gaia} DR2)  we predict astrometric microlensing events by foreground stars of high proper motion passing by a background star in the coming years.}
{We compile a list of approximately 148,000 high-proper-motion stars within {\it Gaia} DR2 with \mbox{\(\mu_{tot}> 150\,\mathrm{mas/yr}\)}. We then search for background stars  close to their paths and calculate the dates and separations of the closest approaches.  Using color and absolute magnitude, we determine  approximate masses of the lenses. Finally, we calculate the expected astrometric shifts and magnifications of the predicted events.}
   {We  detect two ongoing microlensing events by the high-proper-motion stars  \mbox{Luyten~143-23} and Ross 322 and predict closest separations of  \((108.5 \pm 1.4)\,\mathrm{mas}\) in July 2018 and \((125.3\pm 3.4) \, \mathrm{mas}\) in August 2018, respectively.
The respective expected  astrometric shifts are  \((1.74 \pm 0.12)\, \mathrm{mas} \) and \( (0.76 \pm 0.06)\, \mathrm{mas} \).    Furthermore, \mbox{Luyten~143-23} will  pass by another star in March 2021 with a  closest separation of \( (280.1\pm 1.1)\, \mathrm{mas}\), which results in an expected shift of \( (0.69\pm 0.05)\, \mathrm{mas}\). }
   {}


   \maketitle
%

\section{Introduction}

Gravitational lensing is a powerful tool to probe our universe on very different mass and distance scales, from exoplanets in the Milky Way  to multiply imaged quasars and giant arcs at cosmological distance scales.
When a foreground star (``lens'') passes a background star (``source''), two effects can be observed due to gravitational lensing: A magnification of the source (photometric microlensing), and a shift of the source position (astrometric microlensing). Whereas more than 10,000 photometric microlensing events have been observed ‚Äì including a few dozen extrasolar planet detections \citep[e.g.][]{2015AcA....65....1U},  astrometric microlensing was detected for the first time only recently  \citep[][]{2017ApJ...843..145K}.  
Due to its dependence on mass,  astrometric microlensing is one of the few possibilities to ``weigh'' stellar objects   with a precision of about one percent \citep{1995AcA....45..345P}. 

For photometric microlensing, the impact parameter between lens and source needs to be of the order of the angular Einstein radius or smaller  (\(\Delta \theta \lesssim 1\, \theta_{E}\)) for measurable effects. 
Astrometric microlensing  can be detected at much larger angular separations between lens and source  (i.e., \(\Delta \theta \gg 1 \, \theta_{E}\)). Furthermore, it is possible to {\it predict} astrometric microlensing events from precise proper motions and positions of  lens and source \citep{1995AcA....45..345P}. 
For the  prediction of astrometric events,  in particular faint nearby stars with high proper motions are of interest. High proper motions are preferred because the covered area within a given time is larger, and therefore microlensing events are more likely. Furthermore, a high proper motion leads to a significant positional shift on a shorter timescale. Nearby stars are preferred, because their Einstein radius is larger and therefore the expected shift is also larger. Furthermore,  faint lenses are favourable,  since the measurement of the source position is less contaminated by the lens. 
Astrometric microlensing events have already been predicted by, for example, \cite{2011A&A...536A..50P}, \cite{2014ApJ...782...89S} and \citet{2018arXiv180511638M}. 
 The second data release of {\it Gaia} \citep[{\it Gaia} DR2,][]{2018arXiv180409365G} now  provides a highly improved data set for such studies. Not only are the proper motions of the lenses  provided by {\it Gaia} DR2, but  also  precise parallaxes which are needed to calculate the mass of the lens afterwards, as well as  the proper motion of the source. These large improvements in data quality and quantity lead to much more precise predictions. In addition, the excellent resolution and accuracy of  {\it Gaia}  provides the opportunity to actually measure the predicted shifts of the background sources. The events during the {\it Gaia} Mission are therefore of particularly high scientific interest.  

This letter  is structured as follows: In Section \ref{chapter:methode} we explain the basics of our method, in Section \ref{chapter:results} we present the predicted microlensing events of two nearby stars, and in Section \ref{chapter:conclusion}  we summarise our results and present conclusions.

\section{Method}
 \label{chapter:methode} 

We use a method similar to that of \cite{2011A&A...536A..50P} in order to search for  new astrometric microlensing events.
Here we briefly delineate the main steps of our method; the complete method will be detailed in an upcoming paper (Kl\"{u}ter et al. in prep.). 

We start with a list of  high-proper-motion stars (\(\mu_{tot}>150\,\mathrm{mas/yr}\))  from {\it Gaia} DR2 and apply  the following quality cuts, using parallax (\(\varpi\)), proper motion (\(\mu\)) and two Gaia specific columns, to eliminate possibly erroneous data: \(\varpi > 8  \sigma_{\varpi}\), \(\varpi/\mu_{tot} < 0.3\,\mathrm{yr}\) and \mbox{\(\texttt{phot\_g\_n\_obs}^{2}\cdot \texttt{phot\_g\_mean\_flux\_over\_error} > 10^{6}\)}

Approximately 148,000 stars fulfill these criteria. For those potential lenses, we search for background stars in a rectangular box defined by the J2015.5 positions and the positions in 50~years, with a box width of \(\pm7''\) perpendicular to the direction of the proper motion. In the following, the combination of a high-proper-motion foreground lens and a background source is called ``candidate''. The source parameters are labelled with the prefix   ``\(\text{Sou}\_\)''.
To avoid binary stars in our  candidate list, we exclude common proper motion pairs.
Using the {\it Gaia} DR2 positions, proper motions, and parallaxes, we  then determine the minimum separation between source and lens, and the epoch of  closest approach for roughly 68,000 candidates. 

In order to obtain a realistic value for the expected astrometric shifts of our candidates,  we estimate an approximate mass for the lens in the following way.
We sort the potential lenses into the  three classes; white dwarfs (WD), main sequence (MS) stars, and red giants (RG), by using the following cuts:
\begin{equation}
\begin{aligned}
WD:&\qquad G_{BP, abs} \ge 4\cdot (G-G_{RP})^{2}+4.5\cdot(G - G_{RP}) +6\\
RG:&\qquad G_{BP, abs} \le -3 \cdot (G-G_{RP})^{2}+8 \cdot (G-G_{RP})  - 1.3,
\end{aligned}
\end{equation}
where \(G_{BP, abs}\) is the absolute \(G_{Bp}\) magnitude.
For white dwarfs   and  red giants,   we use  approximate masses of \( (0.65\pm0.15)\,M_{\odot}\) and \( (1.0\pm 0.5) \, M_{\odot}\), respectively.
For MS stars we use the following relation with absolute G magnitude (\(G_{abs}\)), determined from a list  of  temperatures and stellar radii for different stellar types \citep{2013ApJS..208....9P}, the translation between {\it Gaia} and Johnson filters \citep{2010A&A...523A..48J}, and mass-luminosity relations \citep{2005essp.book.....S}: 
\begin{equation}
\begin{aligned}
(G_{abs} < 8.85):\\ 
\log\left(\frac{M}{M_{\odot}}\right)   &= 0.00786\,G_{abs}^{2} -0.290\,G_{abs} + 1.18\\
(8.85 <  G_{abs} < 15):\\       
\log\left(\frac{M}{M_{\odot}}\right)& =         -0.301\,G_{abs}  + 1.89\\
(15 < G_{abs}):\\
M&= 0.07 M_{\odot},
\label{Eq:Gmag}        
\end{aligned}
\end{equation} 
The first two relations are due to different slopes in the mass-luminosity relations, and the third one is for brown dwarfs. For MS stars, we consider an error of \(10\%\).
We also use these equations when {\it Gaia} DR2 does not provide colour information for the lens, that is, we assume those stars are MS stars.  
As the next step, we calculate the angular Einstein radii from the masses and the {\it Gaia} DR2 parallaxes:
\begin{equation}
\theta_{E} = 2.854\,\mathrm{mas} \sqrt{\frac{M}{M_{\odot}}\cdot\frac{\varpi-\text{Sou}\_\varpi}{1\,\mathrm{mas}}}
\label{equation:einstein_radius}
.\end{equation}
Finally,  we determine the expected shift of the centre of light of the two images  \citep{1998ApJ...494L..23P},
\begin{equation}
\delta{\theta_{c}} = \frac{{u}}{u^{2}+2} \cdot{\theta_{E}},
\label{equation:shift}
\end{equation} 
where u is the unlensed angular separation between lens and source in units of \(\theta_{E}\).
We note that for large separations (u > 5),   Eq.~\ref{equation:shift} is also a good approximation for the shift of the brighter image only. For small separations, luminous-lens effects have to be taken into account, that is, a factor of \(A/(A+f_{LS})\) has to be added, where \(A\) is the magnification of the event, and \(f_{LS}\) is the flux ratio between lens and source.

\section{Astrometric microlensing event}
 \label{chapter:results}
We found and report here two currently ongoing astrometric microlensing events by the high-proper-motion stars  \mbox{Luyten~143-23}  and Ross 322.
The expected maximum shifts are \mbox{\( (1.74 \pm 0.12)\, \mathrm{mas}\)} and 
\( (0.76\pm 0.06) \, \mathrm{mas} \), respectively.

\begin{figure}[]
\center
\includegraphics[width=7.6cm]{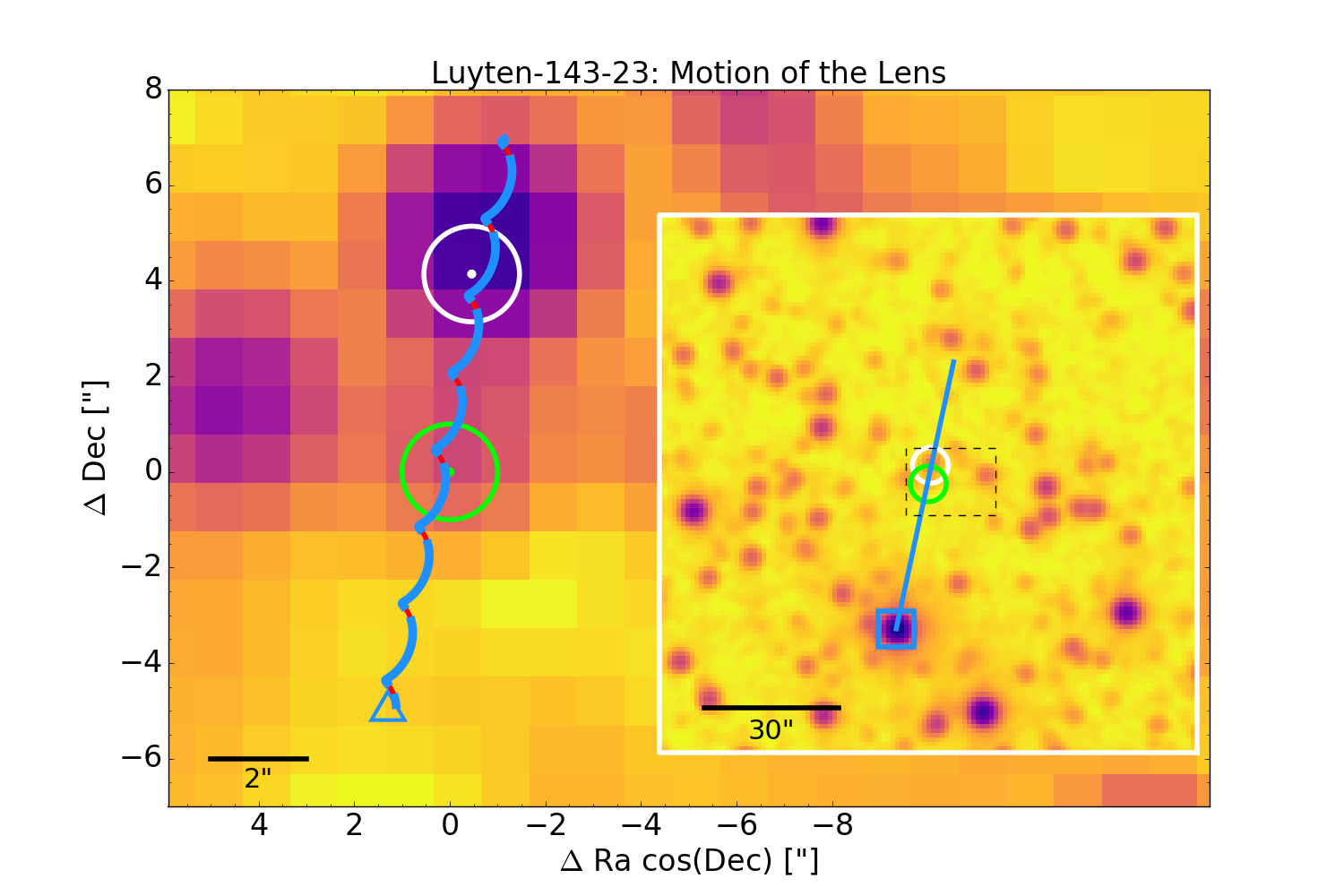}
\includegraphics[width=7.6cm]{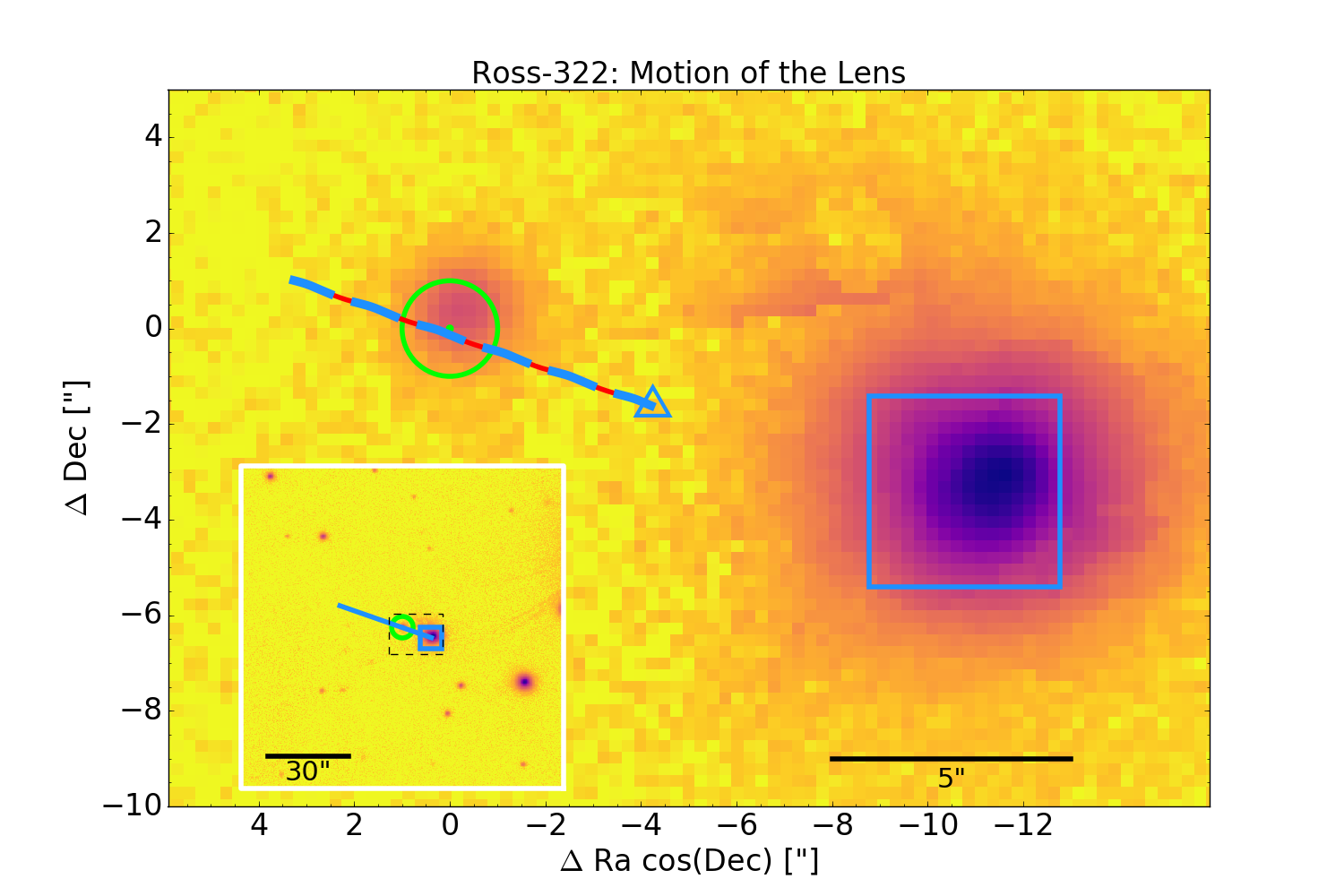}
\caption{The top panel shows a 2MASS image with the two background stars that Luyten 143-23 passes by in July 2018 (green circle) and in March 2021 (white circle), respectively. The bottom panel shows a \mbox{Pan-STARRS} image with the background star which Ross 322 passes by in August 2018 (green circle). The exact positions of the background sources are indicated with small dots at the centres of the circles. Their motions and the uncertainties are smaller than the size of the dots. The triangles indicate the J2015.5 positions of the lens stars measured by Gaia. The predicted motions are shown as thick blue/thin red curves, where the thin red parts indicate the epochs at which the stars cannot be observed. In both panels, the original positions (at earlier epochs) of the lens stars in the 2MASS and PanSTARRS images are indicated as squares.  Larger views of the sky areas are displayed in the insets. The blue lines in the insets show the motion of the lens stars until 2035.}
\label{Figure:2MASS}
\end{figure}

\subsection{\mbox{Luyten~143-23}}
\mbox{Luyten~143-23}
is an M dwarf  (M4V) with a parallax of  \(206.8\, \mathrm{mas}\) and an absolute proper motion of  \(1647.2\,\mathrm{mas/yr}\). Its apparent G magnitude is  \(11.9\, \mathrm{mag}\)  (see Table~\ref{tab:lens} for more details). By using our  \(G_{abs} - \mathrm{Mass}\)  relation (Eq. \ref{Eq:Gmag}) we determine an approximate mass of \( 0.12\,M_{\odot}\).
We found that it currently  passes a \(G = 18.5\, \mathrm{mag}\) star with a closest approach on  July 7,  2018  and a smallest separation of \(108.53\,\mathrm{mas}\). Position, proper motion, parallax and magnitudes for the background source star are given in the top part of Table \ref{tab:events},  information on the closest approach can be found in the bottom part of Table \ref{tab:events}. 
The top panel of Fig.~\ref{Figure:2MASS}  shows a 2MASS \citep[][]{2006AJ....131.1163S} image of \mbox{Luyten~143-23}; the interesting background star is marked with a green circle. The motion of the background star within 5 years and the  positional uncertainties are  smaller than the centered dot.  The predicted path of \mbox{Luyten~143-23} is shown as a thick blue/thin red line, where the red parts indicate the times  at which \mbox{Luyten~143-23} is not observable. The blue  triangle indicates its J2015.5 position listed by {\it Gaia} DR2.
We note that the observability is only a rough estimate, since it depends on the location of the observatory and the capability of the instruments. 

Due to the  large Einstein radius of \( \theta_E = 14.0\,\mathrm{mas}\)  caused by the small distance of \mbox{Luyten~143-23}, a maximum shift of \mbox{\( (1.74 \pm 0.12)\, \mathrm{mas}\)} is expected. The unlensed motion of the background star as well as the predicted path due to  lensing are shown in the top panel of  Fig.~\ref{Figure:BGS}. 
Finally, the absolute shift as a function of time is displayed in the top panel of Fig.~\ref{Figure:Shift}. 
Due to the high proper motion of \mbox{Luyten~143-23,} we expect a rapid change of the astrometric shift  between June and September 2018 (\(1.3\,\mathrm{mas}\) in 3 months). Unfortunately, \mbox{Luyten~143-23} is not observable in August/September 2018 from the ground or by Gaia. 

We  found a further close encounter of \mbox{Luyten~143-23}  with a \(G = 17.0\, \mathrm{mag}\) star to occur in  2021.   The smallest separation of \( (280.1\pm 1.1) \,\mathrm{mas}\)  is reached on  March 9, 2021;  we expect a maximum shift of  \( (0.69\pm 0.05) \,\mathrm{mas}\). 
This background star is marked with a  white circle in Fig. \ref{Figure:2MASS}. The expected motion and the astrometric shift are shown in the middle panels of \mbox{Figs. \ref{Figure:BGS} and \ref{Figure:Shift}}.

\subsection{Ross 322}

\begin{figure}[]
\center

\includegraphics[width=8.6cm]{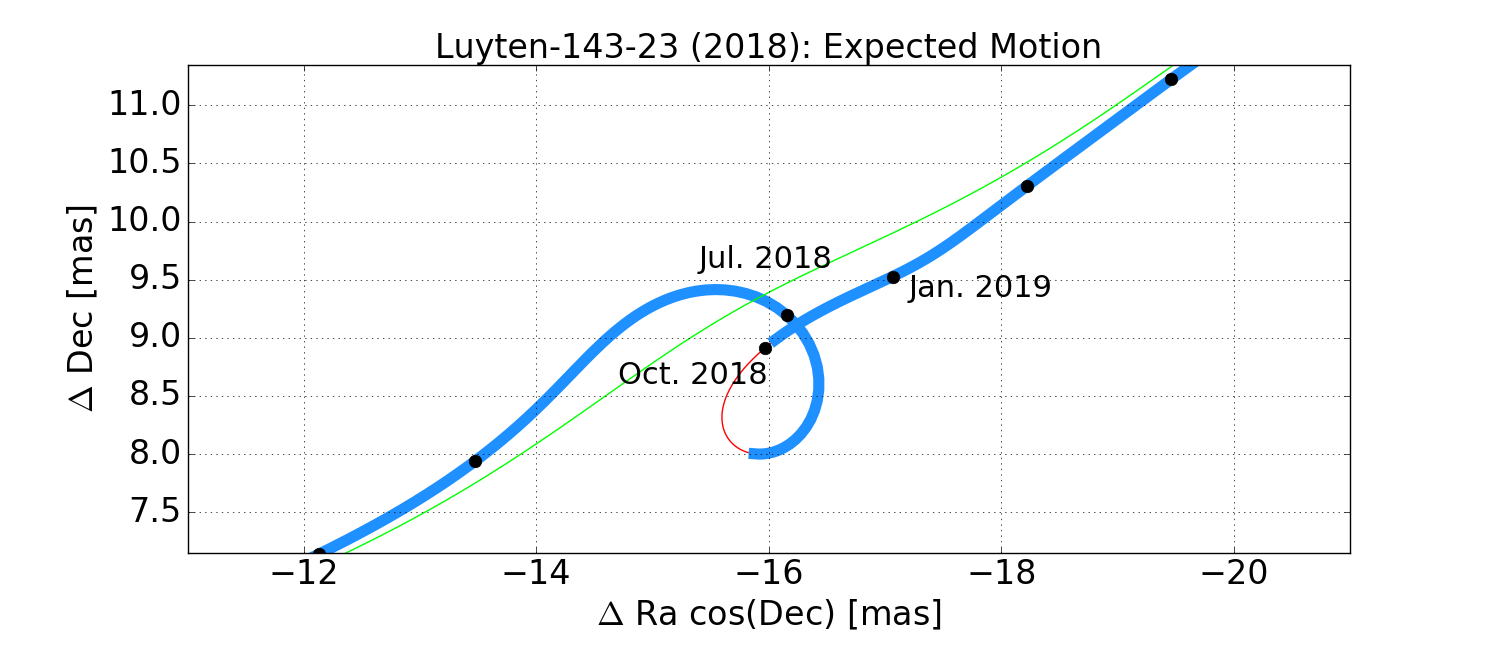}
\includegraphics[width=8.6cm]{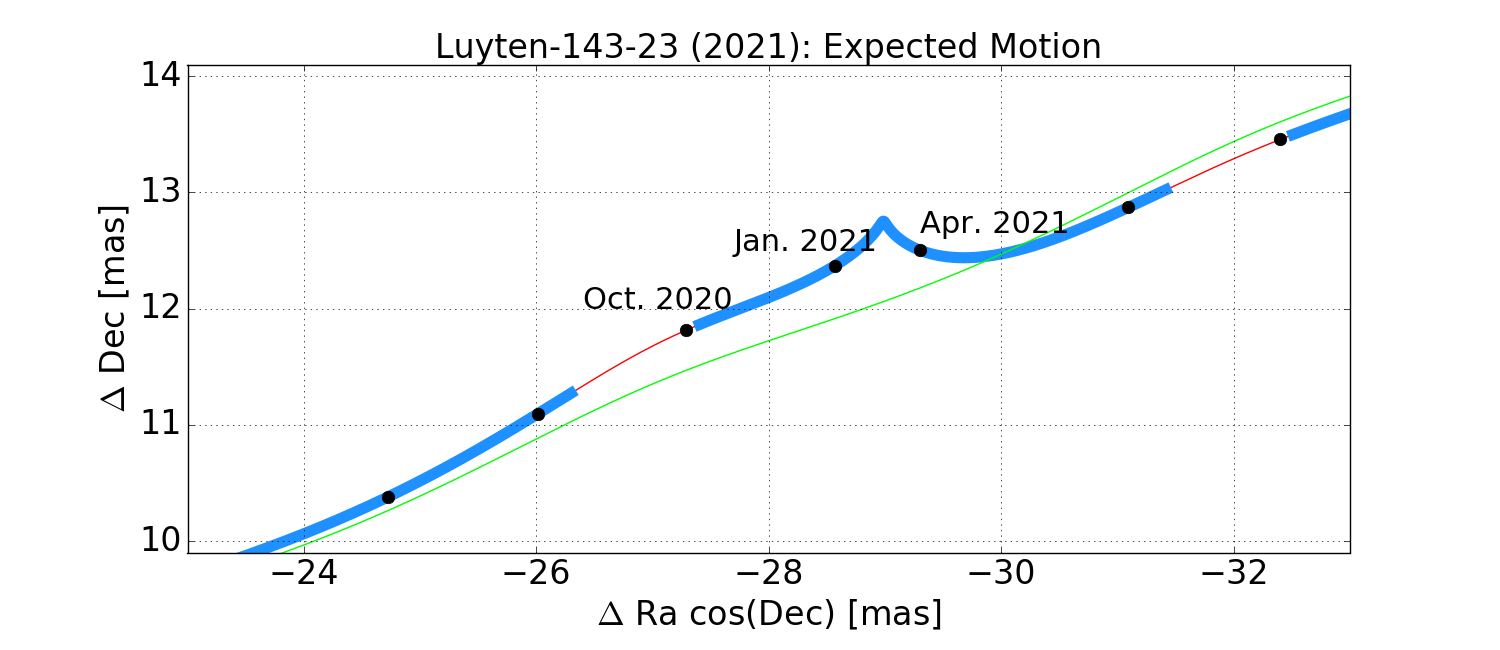}
\includegraphics[width=8.6cm]{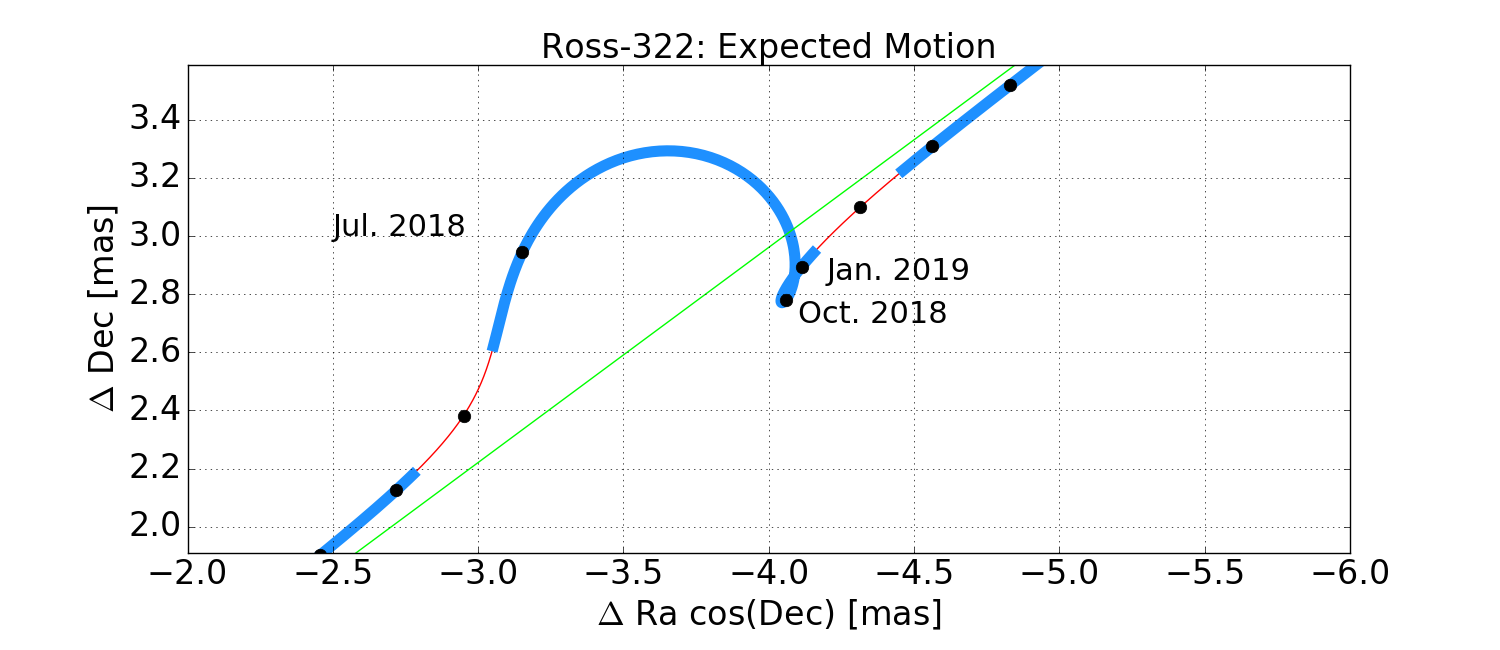}

\caption{Predicted motions of the background stars for the events by \mbox{Luyten~143-23}  in 2018 (Top), in 2021 (Middle),  and by Ross 322  in 2018 (Bottom). The origin of the coordinate system is the background star's J2015.5 position. The thin green lines show the unaffected motions of the background stars, the expected paths are shown as thick blue/thin red lines (red parts indicate the epochs at which the star is not observable). \(\delta\theta_c\) is the difference between both lines at the same epoch.The black dots indicate
equal time intervals of 3 months.} 
\label{Figure:BGS}
\end{figure}

\begin{figure}[]
\center
\includegraphics[width=8.6cm]{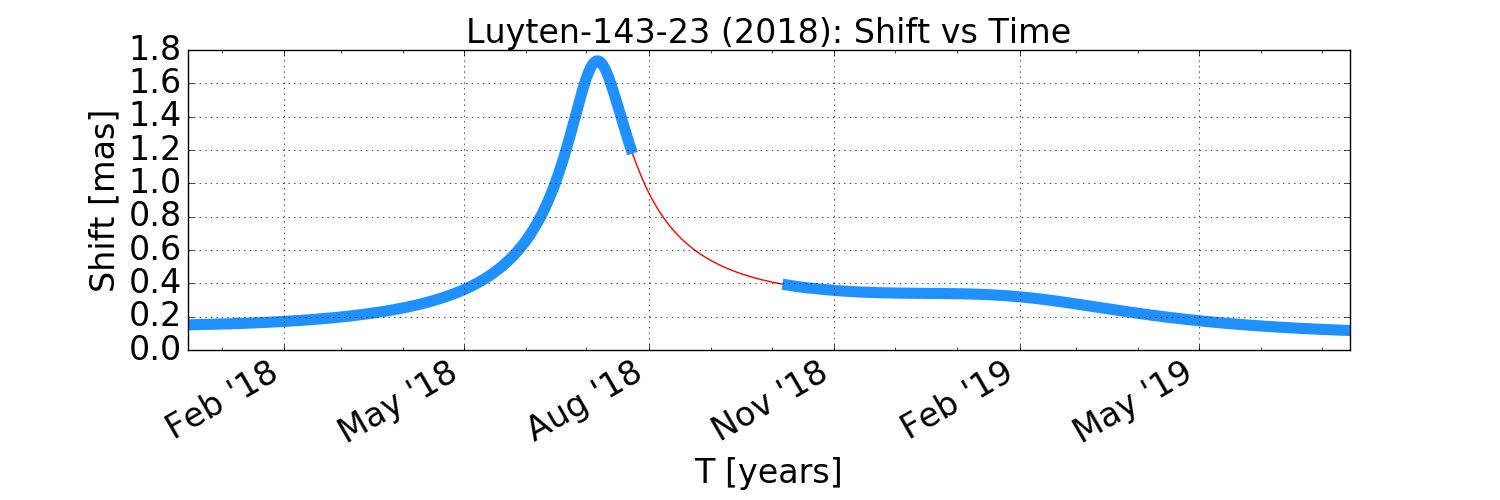}
\includegraphics[width=8.6cm]{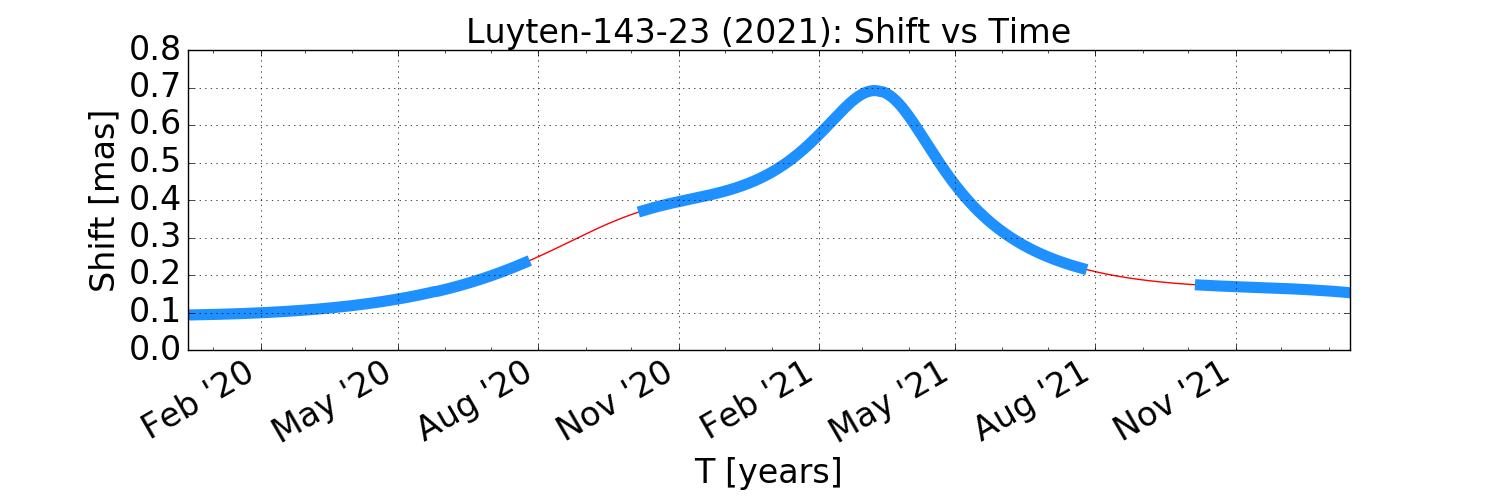}
\includegraphics[width=8.6cm]{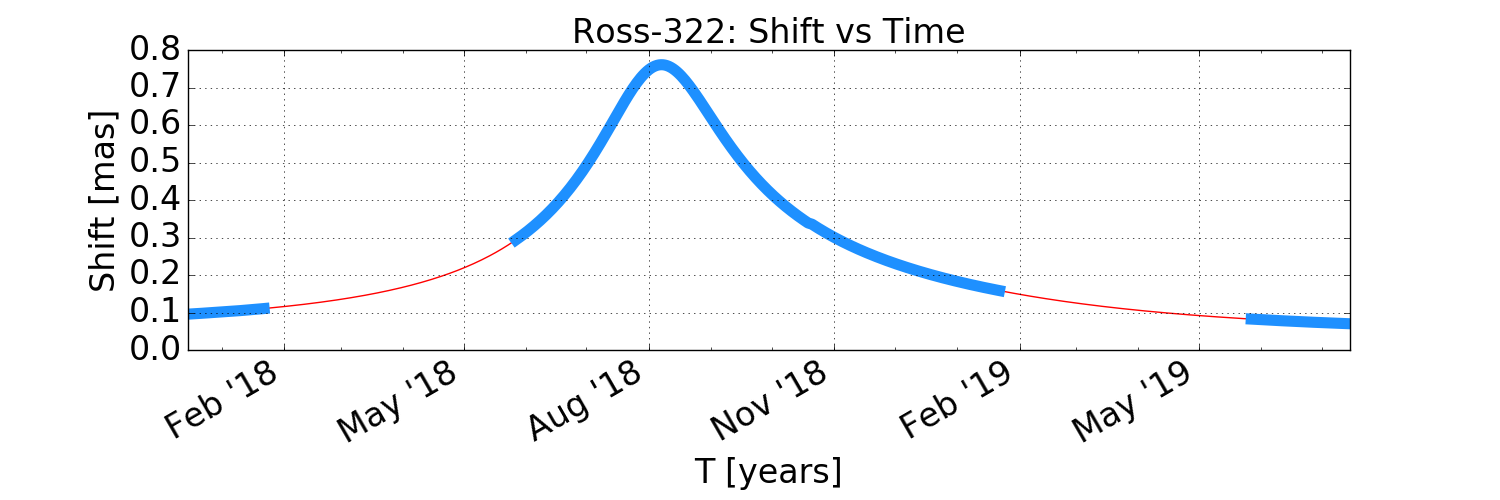}

\caption{Expected  shifts as a function of time for events by \mbox{Luyten~143-23}  in 2018 (Top), in 2021 (Middle),  and by Ross 322 in 2018  (Bottom). The thin red parts indicate the epochs at which the star is not observable.}
\label{Figure:Shift}
\end{figure}

Ross 322 is an M dwarf (M2V) with a parallax of \(42.5\,\mathrm{mas}\).
Its absolute proper motion is \(1455.2\,\mathrm{mas/yr}\) (Tab. \ref{tab:lens}). We determine an approximate mass of \(0.28 \, M_{\odot}\). 
Ross 322 currently passes by a \(G = 18.6\,\mathrm{mag}\) star with a closest approach on  August 8, 2018,  with a separation of \( (125.3 \pm 3.4) \,\mathrm{mas}\). Since the Einstein radius of  Ross 322 is smaller than that of \mbox{Luyten~143-23,} the expected shift is only \( (0.76 \pm 0.06) \,\mathrm{mas}\). However, this should still be measurable 
with sufficient accuracy.
The important parameters are also displayed in Table \ref{tab:events} and the same plots as for \mbox{Luyten~143-23} are given in the bottom panels of  Figs. \ref{Figure:2MASS} to \ref{Figure:Shift}, superposed on a  Pan-STARRS image \citep[][]{2018AAS...23110201C}.

\begin{table}[]
\center
\caption{\label{tab:lens} Important {\it Gaia} DR2  parameters for the lenses.  
 Listed are source\,ID (\(ID\)), J2015.5 position  (\(Ra\), \(Dec\)), proper motion (\(\mu\)), parallax (\(\varpi\)), and magnitude (\(G\), \(G_{RP}\), \(G_{BP}\)) for the lens stars.
Further information for the lens stars can be extracted from {\it Gaia} DR2 using the Source\,ID. }  

\begin{tabular}{|l|rr|}
\hline
\multicolumn{1}{|c}{}& \multicolumn{1}{c}{\mbox{Luyten~143-23}} & \multicolumn{1}{c|}{Ross 322}\\
\hline
\hline

\(ID\)&\(5254061535097566848\)&\(314922605759778048\)\\
\hline
\(Ra\)&\(161.085375416^\circ\)&\(16.956492031^\circ\)\\
\(Dec\)&\(-61.202862014^\circ\)&\(34.210553416^\circ\)\\
\(\sigma_{Ra}\)&\(0.10\,\mathrm{mas}\)& \(0.05\,\mathrm{mas}\)\\
\(\sigma_{Dec}\)&\(0.13\,\mathrm{mas}\)& \(0.04\,\mathrm{mas}\)\\%
\hline
\(\mu_{Ra}\)&\(-346.43\,\mathrm{mas/yr}\)& \(1373.67\,\mathrm{mas/yr}\)\\
\(\mu_{Dec}\)&\(1610.34\,\mathrm{mas/yr}\)& \(-480.33\,\mathrm{mas/yr}\)\\
\(\sigma_{\mu_{Ra}}\)&\(0.17\,\mathrm{mas/yr}\)& \(0.09\,\mathrm{mas/yr}\)\\
\(\sigma_{\mu_{Dec}}\)&\(0.25\,\mathrm{mas/yr}\)& \(0.08\,\mathrm{mas/yr}\)\\
\hline
\(\varpi\)&\(206.82\,\mathrm{mas}\)& \(42.51\,\mathrm{mas}\)\\
\(\sigma_{\varpi}\)&\(0.08\,\mathrm{mas}\)& \(0.05\,\mathrm{mas}\)\\
\hline
\(G \)&\(11.9\,\mathrm{mag}\)& \(12.4\,\mathrm{mag}\)\\
\(G_{RP} \)&\(10.9\,\mathrm{mag}\)& \(11.3\,\mathrm{mag}\)\\
\(G_{BP} \)&\(14.2\,\mathrm{mag}\)& \(13.6\,\mathrm{mag}\)\\
\hline
\end{tabular}
\end{table}

\begin{table*}[]
\center

\caption{\label{tab:events} Important parameters for the astrometric microlensing events reported here. 
 The table lists  source\,ID (\(\text{Sou}\_ID\)), J2015.5 position (\(\text{Sou}\_Ra\), \(\text{Sou}\_Dec\)), proper motion (\(\text{Sou}\_\mu\)) parallax(\(\text{Sou}\_\varpi\)), and G magnitude (\(\text{Sou}\_G\), \(\text{Sou}\_G_{RP}\), \(\text{Sou}\_G_{BP}\)) for the background stars. Furthermore, the angular Einstein radii (\(\theta_{E}\)),  epoch and distance of  closest approach (\(T_{min}\),\,\(D_{min}\)), and the maximum expected shifts (\(\Delta\theta_{max}\)) are shown. Further information for the background stars can be extracted from {\it Gaia} DR2 using the Source\,ID. }
\begin{tabular}{|l|rrr|}

\hline
\multicolumn{1}{|c}{}& \multicolumn{1}{c}{\mbox{Luyten~143-23}  (2018)}& \multicolumn{1}{c}{\mbox{Luyten~143-23} (2021)}  & \multicolumn{1}{c|}{Ross 322 (2018)}\\
\hline
\hline
\(\text{Sou}\_ID\)&\(5254061535052907008\)&\(5254061535097574016\)&\(314922601464808064\)\\
\hline
\(\text{Sou}\_Ra\)&\(161.08462829^\circ\)&\(161.08436459^\circ\)&\(16.95791867^\circ\)\\
\(\text{Sou}\_Dec\)&\(-61.20147495^\circ\)&\(-61.20032468^\circ\)&\(+34.21100433^\circ\)\\
\(\text{Sou}\_\sigma_{Ra}\)&\(0.16\,\mathrm{mas}\)&\(0.06\,\mathrm{mas}\)& \(0.27\,\mathrm{mas}\)\\
\(\text{Sou}\_\sigma_{Dec}\)&\(0.16\,\mathrm{mas}\)&\(0.06\,\mathrm{mas}\)& \(0.38\,\mathrm{mas}\)\\%
\hline
\(\text{Sou}\_\mu_{Ra}\)&\(-4.85\,\mathrm{mas/yr}\)&\(-5.20\,\mathrm{mas/yr}\)& \(-1.12\,\mathrm{mas/yr}\)\\
\(\text{Sou}\_\mu_{Dec}\)&\(2.82\,\mathrm{mas/yr}\)&\(2.17\,\mathrm{mas/yr}\)& \(0.83\,\mathrm{mas/yr}\)\\
\(\text{Sou}\_\sigma_{\mu_{Ra}}\)&\(0.38\,\mathrm{mas/yr}\)&\(0.13\,\mathrm{mas/yr}\)& \(0.60\,\mathrm{mas/yr}\)\\
\(\text{Sou}\_\sigma_{\mu_{Dec}}\)&\(0.33\,\mathrm{mas/yr}\)&\(0.12\,\mathrm{mas/yr}\)& \(0.62\,\mathrm{mas/yr}\)\\
\hline
\(\text{Sou}\_\varpi\)&\(0.09\,\mathrm{mas}\)&\(0.07\, \mathrm{mas}\)& \(-0.05\,\mathrm{mas}\)\\
\(\text{Sou}\_\sigma_{\varpi}\)&\(0.19\,\mathrm{mas}\)&\(0.07\,\mathrm{mas}\)& \(0.44\,\mathrm{mas}\)\\
\hline
\(\text{Sou}\_G \)&\(18.5\,\mathrm{mag}\)&\(17.0\, \mathrm{mag}\)& \(18.6\,\mathrm{mag}\)\\
\(\text{Sou}\_G_{RP} \)&\(16.9\,\mathrm{mag}\)&\(15.9\, \mathrm{mag}\)& \(17.6\,\mathrm{mag}\)\\
\(\text{Sou}\_G_{BP} \)&\(19.1\,\mathrm{mag}\)&\(18.0\, \mathrm{mag}\)& \(18.6\,\mathrm{mag}\)\\
\hline
\hline
\(\theta_{E}\)&\( 14.0 \,\mathrm{mas}\)&\(14.0\,\mathrm{mas}\)& \(9.8\,\mathrm{mas}\)\\
\(\sigma{\theta_{E}}\)&\(0.7\,\mathrm{mas}\)&\(0.7\,\mathrm{mas}\)& \(0.5  \,\mathrm{mas}\)\\
\(T_{min}\,(TCB) \)&\( J2018.51236 \,\mathrm{yr}\)&\(J2021.18674\,\mathrm{yr}\)& \(J2018.5995 \,\mathrm{yr}\)\\
\(\sigma{T_{min}}\)&\(0.00051\,\mathrm{yr}\)&\(0.00063\,\mathrm{yr}\)& \(0.0022\,\mathrm{yr}\)\\
\(D_{min}\)&\( 108.5 \,\mathrm{mas}\)&\(280.1\,\mathrm{mas}\)& \(125.3\,\mathrm{mas}\)\\
\(\sigma{D_{min}}\)&\(1.4  \,\mathrm{mas}\)&\(1.1\,\mathrm{mas}\)& \(3.4\,\mathrm{mas}\)\\
\(\Delta\theta_{max}\)&\(  1.74\,\mathrm{mas}\)&\(0.69\,\mathrm{mas}\)& \(  0.76\,\mathrm{mas}\)\\
\(\sigma{\Delta\theta_{max}}\)&\(0.12\,\mathrm{mas}\)&\(0.05\,\mathrm{mas}\)& \(0.06\,\mathrm{mas}\)\\
\hline
\end{tabular}
\end{table*}

\section{Summary and Conclusions}
\label{chapter:conclusion}
We report two ongoing astrometric microlensing events by the two nearby stars \mbox{Luyten~143-23} and Ross 322. They reach their closest approaches  in July 2018 and  August 2018, respectively. Thanks to the precise data of {\it Gaia} DR2, we were able to predict the separations  between foreground and background stars as a function of time with an accuracy of a few percent and the time of the closest approach with an accuracy of a few hours.
Because of the large Einstein radii (\(14.0\,\mathrm{mas}\); \(9.8\,\mathrm{mas}\)), we expect  a measurable effect even though the minimum separation is above \(100\, \mathrm{mas}\). 
\mbox{Luyten~143-23} will pass a further star in March 2021 with a closest separation of \(280.1\, \mathrm{mas}\) and an expected shift of \(0.69\, \mathrm{mas}\).  
For these separations,  it is also possible to resolve both stars with high-resolution instruments such as GRAVITY \citep{2017A&A...602A..94G}, although the background source is approximately five magnitudes fainter (K band) than the lens. 
Further, the high proper motions  of the lenses lead to a fast change of the astrometric shifts.
Therefore, both events are ideal for monitoring with high-accuracy  astrometric instruments like {\it Gaia}, GRAVITY, Keck, or LBT.
Due to the precise prediction, observations of both events offer the possibility to determine the masses of the two nearby high-proper-motion stars  \mbox{Luyten~143-23} and Ross 322 with a precision of a few percent. 
An initiative to acquire  observations with GRAVITY and at the Keck Telescope during the two ongoing events has been launched by the authors.

\nocite{2005ASPC..347...29T}

\begin{acknowledgements}

This work has made use of results from the ESA space mission {\it {\it Gaia}}, the data from which were processed by the {\it {\it Gaia}} Data Processing and Analysis Consortium (DPAC). Funding for the DPAC has been provided by national institutions, in particular the institutions participating in the {\it {\it Gaia}} Multilateral Agreement. The {\it {\it Gaia}} mission website is:
http://www.cosmos.esa.int/{\it Gaia}. Some of the authors are members of the {\it  Gaia} Data Processing and Analysis Consortium (DPAC).\\
This publication makes use of data products from the Two Micron All Sky Survey, which is a joint project of the University of Massachusetts and the Infrared Processing and Analysis Center/California Institute of Technology, funded by the National Aeronautics and Space Administration and the National Science Foundation.\\
The Pan-STARRS1 Surveys (PS1) and the PS1 public science archive have been made possible through contributions by the Institute for Astronomy, the University of Hawaii, the Pan-STARRS Project Office, the Max-Planck Society and its participating institutes, the Max Planck Institute for Astronomy, Heidelberg and the Max Planck Institute for Extraterrestrial Physics, Garching, The Johns Hopkins University, Durham University, the University of Edinburgh, the Queen's University Belfast, the Harvard-Smithsonian Center for Astrophysics, the Las Cumbres Souservatory Global Telescope Network Incorporated, the National Central University of Taiwan, the Space Telescope Science Institute, the National Aeronautics and Space Administration under Grant No. NNX08AR22G issued through the Planetary Science Division of the NASA Science Mission Directorate, the National Science Foundation Grant No. AST-1238877, the University of Maryland, Eotvos Lorand University (ELTE), the Los Alamos National Laboratory, and the Gordon and Betty Moore Foundation.\\
This research has made use of the SIMBAD database,
operated at CDS, Strasbourg, France.\\
We gratefully acknowledge the technical support we received from
staff of the e-inf-astro project (BMBF F\"{o}rderkennzeichen 05A17VH2).

\end{acknowledgements}
\bibliographystyle{aa}
\bibliography{Ongoing_Microlensing_Arxive} 

\end{document}